\documentclass{aastex}
\usepackage{emulateapj5}

\newcommand{\kms}{$\,\mbox{km}\,\mbox{s}^{-1}$}

\newcommand{\Msun}{$M_{\odot}$}
\newcommand{\HI}{H{\sc i}\ }

\shortauthors{de Blok \& Walter}
\shorttitle{A giant H\lowercase{{\sc i}} hole in NGC\,6822}

\begin{document}

\title{Evidence for tidal interaction and a supergiant HI shell in the 
Local Group dwarf galaxy NGC 6822}

\author{W.J.G.~de~Blok\altaffilmark{1}}
\affil{Australia Telescope National Facility}
\affil{PO Box 76, Epping NSW 1710, Australia}
\email{edeblok@atnf.csiro.au}
\altaffiltext{1}{Bolton Fellow}

\and

\author{F.~Walter} 
\affil{California Institute of Technology}
\affil{Astronomy Department 105-24,\\ Pasadena, CA 91125, USA}
\email{fw@astro.caltech.edu}
 
\begin{abstract}

We present a wide-field, high spatial and velocity resolution
map of the entire extended H{\sc i} distribution of the Local Group
dwarf galaxy NGC 6822.  The observations were obtained with the
Australia Telescope Compact Array in mosaicing mode. The interstellar
medium of NGC 6822 is shaped by the presence of numerous H{\sc i}
holes and shells, including a supergiant shell, and the effects of
tidal interaction, in the form of a tidal arm and an infalling or
interacting H{\sc i} complex. The H{\sc i} shell is situated outside
the optical galaxy and occupies roughly a quarter of the area of the
main H{\sc i} disk. It measures $2.0\times 1.4$~kpc, making it one of
the largest supergiant H{\sc i} shells ever found.  The giant hole
shows no signs of expansion and no obvious creation mechanism is
evident from our data.  If star formation was the cause, an energy
equivalent of $\sim 100$ supernovae ($10^{53}$\,erg) is needed to
create the hole. We derive an upper limit for the age of order 100
Myr.  The presence of a possible tidal arm indicates that NGC\,6822
may recently have undergone some interaction. An H{\sc i} complex
located in the north-west of the galaxy may be the interaction
partner. We argue that it is likely that these features were created
about 100 Myr ago in an event that also enhanced the star formation
rate.

\end{abstract}


\section{Introduction}

Located at a distance of only $490\pm 40$ kpc (Mateo 1998),
NGC\,6822 is, apart from the LMC/SMC system, the most nearby dwarf
irregular galaxy known. Discovered by Barnard (1884), its strong
resemblance to the LMC/SMC led Perrine (1922) to speculate that
``Barnard's Nebula'' might be an object located outside our
Galaxy. This was confirmed by Hubble (1925) who determined its
distance using Cepheid measurements, making it ``the first object
definitely assigned to a region outside the galactic system.''

NGC\,6822 has no known companions and is not associated with the
concentrations of galaxies surrounding M31 and the Milky Way (see
e.g.\ van den Bergh 1999); rather it belongs to the ``Local Group
Cloud'', an extended cloud of dwarf irregulars (Mateo 1998).  NGC~6822
has a total luminosity of $M_B\!=\!-15.8$ (Hodge et al.\ 1991) and a
total H{\sc i} mass of $1.3 \times 10^8\ M_{\odot}$ (de Blok \& Walter
2000), making it relatively gas-rich. It is a metal poor galaxy, with
an ISM abundance of about 0.2 $Z_{\odot}$ (e.g. Skillman 1989).  It
has a low star formation rate of $\sim 0.06\ M_{\odot}{\rm yr}^{-1}$
(based on H$\alpha$ and FIR fluxes; Mateo 1998, Israel et al.\
1996). In terms of these global properties NGC 6822 is a rather
average and quiescent dwarf irregular galaxy.  In the past little
attention has been paid to the distribution of the neutral ISM of NGC
6822.  This can partly be attributed to the fact that part of the
H{\sc i} emission of NGC\,6822 is at the same heliocentric velocity as
strong foreground emission from our Milky Way.  This ``Galactic
interference'' is hard to remove if only a coarse velocity resolution
is employed, as was the case in the early H{\sc i} studies.  This
confusion, as well as NGC\,6822's large size, may be the reasons why
to date only few \HI studies of this galaxy have appeared in the
literature, even though its inclination, gas-richness and proximity
make it an ideal candidate for studying the rotation curve, dark
matter content and relation between the stellar population and gaseous
ISM.

The earliest \HI study of NGC\,6822 is that by Volders \& H\"ogbom
(1961) who used the 25m Dwingeloo telescope to probe NGC\,6822 and its
environs.  They found evidence for neutral hydrogen rotating in ordered
fashion around the optical center.  Later studies by Davies (1970) and
Roberts (1970), confirmed these results.  Follow-up interferometric
studies by Gottesman \& Weliachew (1977) and Brandenburg \& Skillman
(1998) (see also Hodge et al.\ 1991) showed a structured and
extended ISM in and around the optical galaxy. 

Here, we present the first results of a large project to map the
entire H{\sc i} disk of NGC~6822 in unprecedented detail. We used the
Australia Telescope Compact Array (ATCA) in mosaicing mode and the
Parkes single dish telescope, mapping an area of $1.5 \times1^{\circ}$
($13.5 \times9$\ kpc).  A preliminary inspection of the data showed
that, contrary to expectations (based on the low SFR and its isolated
position in the Local Group), the H{\sc i} appearance is dramatic: it
is dominated by a giant H{\sc i} shell with extended, apparently
tidal, features in the outer disk.  This disturbed H{\sc i} morphology
and how it affects our view of the evolution of NGC 6822 is the
subject of this Letter.

\section{Observations and data reduction}

NGC\,6822 has been observed with the Australia Telescope Compact Array
for $15 \times 12$ hours in its 375, 750D, 1.5A, 6A and 6D
configurations over the period from June 1999 to March 2000. A total
of 8 pointings was observed covering the entire H{\sc i} extent of the
galaxy.  We used a bandwidth of 4 MHz with a channel separation of 0.8
\kms.  Additionally, NGC 6822 was observed with the Parkes single dish
radio telescope using the multibeam receiver in December 1998.  A full
account of the observations will be presented elsewhere (de~Blok \&
Walter, in prep.). In this Letter we restrict ourselves to the medium
resolution data which does not include the 6-km configuration, nor the
single dish data.

The data was reduced and mosaiced together using the {\sc miriad} data
reduction package.  Super-uniform weighting, reducing side lobes in
individual pointings prior to mosaicing, was used. The resulting data
cubes were cleaned with the {\sc miriad mossdi} task.  The clean
synthesized beam of this medium resolution data set measures $89''
\times 24''$ ($222 \times 60$ pc).  The noise per channel
is 8 mJy, resulting in a $5\sigma$ column density sensitivity of $1.6
\times 10^{19}$ cm$^{-2}$.

\section{The H{\sc i} morphology}

Figure 1 shows the integrated H{\sc i} column density map.  The map
was produced by using a two times smoothed and clipped ($2.5\sigma$)
version of the data as a mask.  Also shown is the same map overplotted
on an optical DSS image.  The H{\sc i} is more extended than the
optical distribution, extending out to radii of 40 arcmin ($\sim 6$
kpc). We find a total \HI mass of $1.1 \times 10^8$
\Msun.  A comparison with the single dish Parkes \HI mass of $(1.3 \pm
0.1) \times 10^8$ \Msun\ shows that we are missing very little flux due to
missing short spacings.

The \HI disk contains many holes and shells.  Three striking
features that are the subject of this Letter are marked in Fig.~1: the
hole and arm in the SE, and the cloud in the NW.  

\subsection{The giant H{\sc i} hole} 

In the SE a giant H{\sc i} hole or shell dominates the appearance of
the galaxy.  Its angular size size is $14'\times 10'$ (as indicated in
Fig.~1), corresponding to $2.0\times 1.4$ kpc, measured at a column
density level of $10^{21}$ cm$^{-2}$. We will adopt a mean diameter of
1.7 kpc.  The deviation from a (deprojected) circle can probably be
explained by shear in the outer disk of NGC\,6822 since the rotation
curve flattens at large  radii (de Blok \& Walter, in
prep.).

The giant H{\sc i} shell is hinted at in earlier interferometer maps
by Gottesmann \& Weliachev (1977) of the centre of NGC 6822 and more
clearly mapped in {\sc vla} data presented in Hodge et al.\ (1991) and
Brandenburg \& Skillman (1998).
A major axis cut through the galaxy is shown in Fig.~2. Note that the
inner part of the H{\sc i} hole seems to be completely evacuated.  The
hole does not seem to be expanding.  The dispersion in this area
($\sim 7$~\kms) is significantly lower than in the NW part of the
galaxy ($\sim 9$~\kms) where there are more signs of recent star
formation (Fig.~2, bottom).

\subsection{An H{\sc i} companion?}

It is difficult to tell whether the H{\sc i} complex in the extreme NW
actually belongs to the main disk of NGC\,6822 or whether it is a
companion at a similar heliocentric velocity. The H{\sc i} mass of the
NW complex is $\sim 1.4 \times 10^7\ M_{\odot}$, i.e.\ $\sim 10\%$ of
the total H{\sc i} mass of the total NGC~6822-system.  At the
NGC6822-cloud interface there is a sharp jump in velocity (at $r
\simeq 0.3^{\circ}$ in Fig.~2) which may indicate that it is indeed a
separate system. In principle, the jump could have been created by an
asymmetric blow-out due to star formation. However, the lack of stars,
star-forming regions and the low dispersion (Fig.~2) in that region
make this unlikely.
Furthermore, the NW half contains 20\% more H{\sc i} than the SE half
(a difference of $\sim 1.2\times 10^7\ M_{\odot}$), as measured with
respect to a minor axis passing through the geometrical center.
Assuming that the disk of NGC 6822 is intrinsically symmetric, this
asymmetry can be explained by assuming that the NW cloud (with a mass
of $\sim 1.4\times 10^7\ M_{\odot}$) is a separate system contributing
to the mass in the NW half.

\subsection{Signs of interaction}

A third, ``tidal arm'' feature is visible in the SE. It is unlikely
that it is a conventional spiral arm, due to the absence of star
formation in this part of the galaxy, the absence of any spiral
structure in the optical and the inner H{\sc i} disk, and the overall
asymmetric H{\sc i} morphology.  Whether the material in this arm was
stripped off NGC\,6822's main disk or belonged to an interaction
partner is difficult to tell based on our data. Future numerical
simulations may shed light on this situation.

A search of a $10^{\circ} \times 10^{\circ}$ field surrounding
NGC~6822 using HIPASS data did not yield unknown
\HI companions. One important caveat is, however,
that there is Galactic and HVC emission present in this region of the
sky between $+25$ and $-15$\kms. This velocity range coincides with
that of the SE arm. Possible companions of NGC\,6822 might be hidden
in the strong galactic emission.  
As timescales for an interaction with a known Local Group galaxy are
too large by an order of magnitude, another possibility is that the NW
cloud is the interaction partner. An upper limit to the time scale for
this encounter is of the order of half the rotation period at the
radial distance of the cloud which is $3 \times 10^8$ yr.  A rough
estimate for the timescale can also be derived from the tidal feature
itself: the arm measures some $20'$ or 2.8 kpc. An inspection of the
$pV$ diagram in Fig.~2 suggests a relative velocity between arm and
disk of 10 to 30 \kms. Using 20 \kms\ we derive a kinematic age of
$1.4 \times 10^8$ yr, but any number between 100 and 200 Myr is
probably reasonable.

\section{Origins of the supergiant shell}

In the standard picture, H{\sc i} shells and supershells are caused by
the stellar winds of the most massive stars in a cluster as well as
subsequent SN explosions (for reviews see Kulkarni
\& Heiles 1988, van der Hulst 1996 and Brinks \& Walter 1998).
At first glance, it seems unlikely that this caused the giant shell.
It is located far off the optical center and the bulk of star
formation.  One would need one or more massive stellar clusters at
large radii to create the hole. These clusters ought to have dispersed
over the past 100 Myr, as there is no sign of a young population at
the current epoch.

If the hole was indeed created by star formation, we can derive the
energies and ages involved. Since the hole does not expand any more
and has presumably broken out, it is only possible to make order of
magnitude estimates.  If we assume that the expansion velocity of the
hole has reached values similar to the dispersion of the ambient ISM
($\sim 7$\kms), we derive a kinematic age of 130 Myr. This is an upper
limit for the actual age since the shell was presumably expanding more
rapidly in the past. An age of around 100 million years is therefore
reasonable.  Using Chevalier's equation (Chevalier 1974), we derive an
energy of $10^{53}$\ erg needed to create the H{\sc i} shell,
equivalent to 100 Type II supernovae (using $n_{HI} = 0.1$\
cm$^{-3}$). It is not necessary that these supernovae go off at the
same time, which would need a massive parent cluster. Many sequential
events can create a big hole by superposition. The
kinetic energy of the H{\sc i} shell is $10^{51}$\,erg.

An estimate for the amount of gas removed from the hole $M_h$ can be
made as follows. 

We estimate $z$, the scale height of the disk, following
Puche et al.\ (1992). Using $M_{dyn} = 4.3 \times 10^9\ M_{\odot}$ and
$R_{max} = 5.7$ kpc (de Blok \& Walter, in prep.), we find $z = 0.285$ kpc. 
This yields a
total H{\sc i} mass evacuated from the hole of $M_h = 1.6\times 10^6\
M_{\odot}$. 

The infall of small high-velocity clouds has
sometimes been invoked to explain the largest H{\sc i} supershells in
galaxies (Tenorio-Tagle et al.\ 1988). Observational evidence has been
found in M\,101 (van der Hulst \& Sancisi 1988). However, if the hole
in NGC 6822 were indeed due to infall of a high-velocity cloud we
would expect to see remnants near the hole or at least some
kinematical signature --- neither is obvious (see the position
velocity cut in Fig.~2)
\footnote{In this context, it should be noted that the hole in NGC~6822 has the
same diameter as the hole or shell in NGC 5462, the H{\sc ii} region
in the eastern arm of M101 (Kamphuis, Sancisi \& van der Hulst 1991).}.

\section{The importance of minor interactions}

As briefly noted in the Introduction, based on the luminosity,
gas-richness, metallicity, SFR and other global properties, one would
classify NGC\,6822 as a typical, quiescent dwarf irregular galaxy.
There was thus no reason to suspect that NGC 6822 might have such a
disturbed H{\sc i} disk.  The H{\sc i} data presented here change this
picture completely. They provide evidence for a recent interaction,
which caused a significant increase in star formation, affecting the
morphology of the disk. The results are still around in the form of
the NW H{\sc i} complex, the supergiant H{\sc i} shell and the SE
tidal arm.
The interaction described here was a minor one, as it did not result
in a large starburst and the ejection of large amounts of gas. Dwarf
galaxies can apparently undergo such minor interactions without a
noticeable effect on their {\it global} properties. It is the small
distance to NGC 6822 that enables us to study this process in so much
detail. Low resolution maps made using data from only the 375m ATCA
configuration show that if NGC 6822 had been a factor of $\gtrsim 5$
further away, it would have been impossible to distinguish the NW
complex from the main body, nor would the tidal arm and the hole have
been obvious.

The number of minor interactions in dwarf galaxies may therefore be
larger than one would guess on the basis of low to medium
resolution H{\sc i} observations of more distant galaxies, or on the
basis of observations in other wavelengths.  For example, the low
current SFR in NGC 6822 does not give any indication that the galaxy
was disturbed recently, in contrast with e.g.\ blue compact dwarf
galaxies where the enhanced star formation rates clearly indicate some
disturbance. 
This could lead to a skewed
picture of the importance of minor interactions.

NGC 6822 is one of the very few dwarf systems in the local universe
that allows such a detailed study of its ISM and stellar population.
Although detailed H{\sc i} studies of {\it nearby} dwarf and LSB
galaxies are needed to further investigate the influence of minor
interactions on a more statistical basis (see also e.g.\ Taylor 1997,
Pisano \& Wilcots 1999, Hunter et al.\ 1998), NGC 6822 will remain a
benchmark, as it provides the clearest view we currently have of the
morphology of a dwarf irregular galaxy outside the immediate Milky
Way/LMC/SMC environment.

\subsection{An interaction scenario for NGC 6822}

We now present a possible interaction scenario for NGC\,6822 based on
the observations presented above. As derived earlier, the kinematic
age for the giant shell is $\sim 100$ Myr. The timescale for the NW
complex to interact with NGC 6822 is $\sim 300$ Myr. The kinematical
timescale based on the length of the tidal arm is $\sim 140$ Myr.
Similar timescales have been derived from optical studies: Hodge
(1980) finds evidence for an enhancement in star formation between 75
and 100 Myr ago, while the extensive study by Gallart et al.\ (1996b)
shows that the SFR in NGC 6822 increased by a factor 2 to 6 between
100 and 200 Myr ago. Similarly, Hutchings et al.\ (1999) find evidence
for a 100 Myr young population using WFPC2 imaging.  These are good
indications that something affected the galaxy some 100 to 200 Myr
ago.  This was most likely the passage of the NW complex. This may
have caused the SFR to increase, triggering the formation of the giant
H{\sc i} shell in the tidally disturbed SE part of the galaxy.  The
precise creation mechanism for the large cavity remains a mystery, as
is the case with most of the other supergiant H{\sc i} holes (e.g.\
Walter \& Brinks 1999, Puche et al.\ 1992, Rhode et al.\ 1999).  The
presence of hot gas around the hole would certainly help to rule out
or constrain some of the scenarios discussed here.  In a few other
galaxies there have been successful attempts to locate the heated gas
using X-ray observations (Walter et al.\ 1998). Unfortunately NGC~6822
is located towards an absorbing galactic H{\sc i} column density of
$3\times$10$^{21}$\,cm$^{-2}$ which makes a detection of soft X-ray
emission originating from hot gas virtually impossible even with
today's powerful X-ray telescopes such as Chandra and XMM-Newton.

Follow-up H{\sc i}, optical and infrared observations (allowing
stellar population studies) currently in progress and the added
benefit of very high resolution in the complete H{\sc i} dataset ($6''
= 15$ pc), will will shed more light on the state of the ISM in NGC 6822.

\acknowledgements
FW acknowledges NSF grant AST 96-13717. The Australia Telescope is
funded by the Commonwealth of Australia for operation as a National
Facility managed by CSIRO.  This research has made use of the
NASA/IPAC Extragalactic Database (NED) which is operated by JPL,
Caltech, under contract with NASA and NASA's Astrophysical Data System
Abstract Service (ADS).

\clearpage

\begin{figure}
\epsfxsize=0.5\hsize
\epsfbox{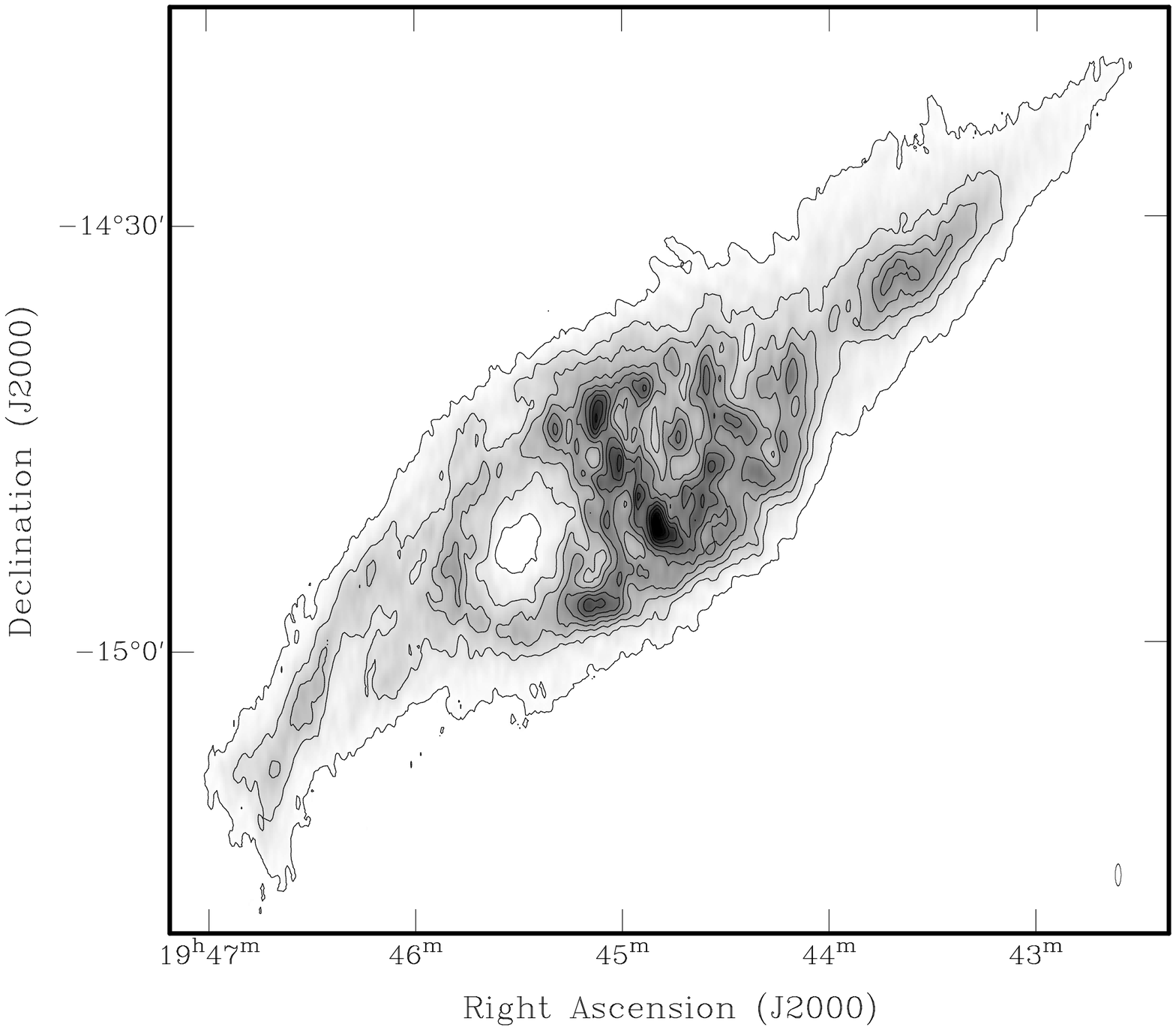}
\epsfxsize=0.5\hsize
\epsfbox{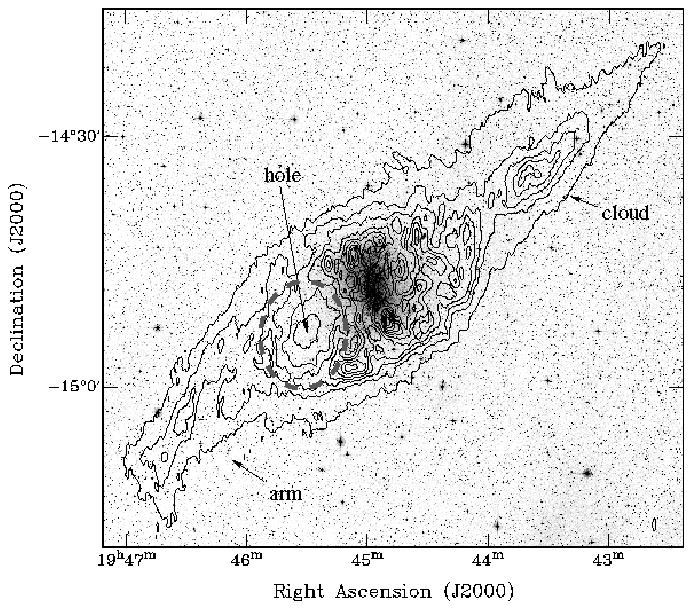}

\figcaption[6822.fig1a.ps,6822.fig1b.ps]{Left panel: integrated H{\sc i} 
column density map. Contours are $(1, 4, 7, 10, ..., 31) \times
10^{20}$ cm$^{-2}$. The beam of $89'' \times 24''$ is indicated in the
lower-right corner. Right panel: Integrated H{\sc i} column density
contours overlaid on optical image from the DSS. Contour values are
same as in left panel. Note how the inner edge of the hole is traced
by the optical emission, notably in the south. The three features of
hole, cloud and arm as discussed in the text are indicated. The
outline of the hole is shown by the grey dashed ellipse.}
\end{figure}

\begin{figure}
\begin{center}
\epsfxsize=0.8\hsize
\epsfbox{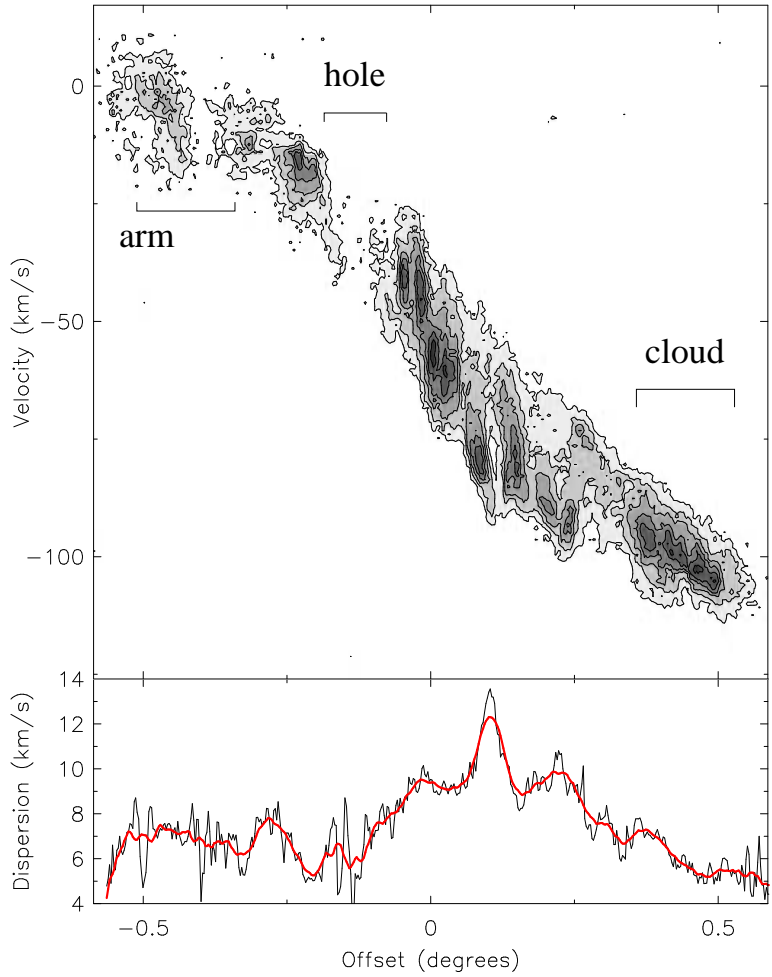}
\figcaption[6822.fig2.ps]{Major axis position velocity diagram, taken at a position
angle of $-54^{\circ}$ through the geometric center of the H{\sc i}
disk. The lowest contour value is $4\sigma$ and levels increase in
intervals of $4\sigma$. Indicated from left to right are the SE arm,
the big hole and the NW cloud. The lower panel shows the velocity
dispersion along the slice. The velocity spacing is 0.8 \kms. The thin
line indicates the local velocity dispersion sampled every $9''$. The
thick line indicates the running mean of the velocity dispersion,
using a boxcar smooth of $2.4'$.  It is clear that the SE part of the
galaxy has a much lower velocity dispersion than the NW part.}
\end{center}
\end{figure}

\end{document}